\documentclass[hidelinks]{llncs}

\usepackage{lineno}
\usepackage{amssymb}
\usepackage{mathtools}
\usepackage{dashbox}
\usepackage{algorithm}
\usepackage{algpseudocode}
\usepackage{setspace}
\usepackage{multirow}
\usepackage[binary-units=true]{siunitx}
\usepackage{subcaption}
\usepackage[nolist]{acronym}
\usepackage[htt]{hyphenat}
\usepackage{longtable}
\usepackage{booktabs}
\usepackage{xcolor}
\usepackage[colorlinks=false]{hyperref}

\usepackage{pifont}
\newcommand{\cmark}{\ding{51}}%
\newcommand{\xmark}{\ding{55}}%

\usepackage{graphicx}
\graphicspath{ {./images/} }

\algnewcommand\algorithmicswitch{\textbf{switch}}
\algnewcommand\algorithmiccase{\textbf{case}}
\algnewcommand\algorithmicassert{\texttt{assert}}
\algnewcommand\Assert[1]{\State \algorithmicassert(#1)}%
\algdef{SE}[SWITCH]{Switch}{EndSwitch}[1]{\algorithmicswitch\ #1\ \algorithmicdo}{\algorithmicend\ \algorithmicswitch}%
\algdef{SE}[CASE]{Case}{EndCase}[1]{\algorithmiccase\ #1}{\algorithmicend\ \algorithmiccase}%
\algtext*{EndSwitch}%
\algtext*{EndCase}%

\definecolor{redbg0}{HTML}{FFD8D5}
\definecolor{redbg1}{HTML}{F3B5AF}
\definecolor{greenbg0}{HTML}{C1F2D1}
\definecolor{greenbg1}{HTML}{AEE5BE}

\usepackage{soul}
\newcommand{\ignore}[1]{}

\newcommand{\til}[1]{\overset{\sim}{#1}}

\usepackage{tikz}
\usetikzlibrary{%
  arrows,
  arrows.meta,
  backgrounds,
  calc,
  fit,
  graphs,
  positioning
}
\tikzstyle{every node}=[node distance=2.25cm]

\usepackage{listings}

\newcolumntype{?}{!{\vrule width 1.25pt}}

\begin{document}

\title{Minimally Comparing Relational Abstract Domains}

\author{%
  Kenny Ballou \orcidID{0000-0002-6032-474X} \and
  Elena Sherman \orcidID{0000-0003-4522-9725}
}
\institute{%
  Boise State University
  \email{kennyballou@u.boisestate.edu,elenasherman@boisestate.edu}
}
\authorrunning{Ballou, et al.}

\begin{acronym}
  \acro{AI}{abstract interpretation}
  \acro{CFG}{control flow graph}
  \acro{DFA}{data-flow analysis}
  \acro{DFS}{depth-first search}
  \acro{DBM}{difference bounded matrix}
  \acrodefplural{DBM}{difference bounded matrices}
  \acro{JVM}{Java virtual machine}
  \acro{LIA}{linear integer arithmetic}
  \acro{NIA}{non-linear integer arithmetic}
  \acro{TVPI}{two variables per inequality}
  \acro{SMT}{satisfiability modulo theories}
  \acro{SMT-LIB}{satisfiability modulo theories library}
  \acro{StInG}{Stanford Invariant Generator}

  \acro{FS}{full state}
  \acro{FG}{Full Graph}
  \acro{CC}{Connect Components}
  \acro{NN}{Node Neighbors}
  \acro{MN}{Minimal Neighbors}
\end{acronym}



\maketitle{}


\begin{abstract}
  Value-based static analysis techniques express computed program invariants as
  logical formula over program variables.  Researchers and practitioners use
  these invariants to aid in software engineering and verification tasks.  When
  selecting abstract domains, practitioners weigh the cost of a domain against
  its expressiveness.  However, an abstract domain's expressiveness tends to be
  stated in absolute terms; either mathematically via the sub-polyhedra the
  domain is capable of describing, empirically using a set of known properties
  to verify, or empirically via logical entailment using the entire invariant
  of the domain at each program point.  Due to \emph{carry-over} effects,
  however, the last technique can be problematic because it tends to provide a
  simplistic and imprecise comparisons.

  We address limitations of comparing, in general, abstract domains via logical
  entailment in this work.  We provide a fixed-point algorithm for including
  the minimally necessary variables from each domain into the compared formula.
  Furthermore, we empirically evaluate our algorithm, comparing different
  techniques of widening over the Zones domain and comparing Zones to an
  incomparable Relational Predicates domain.  Our empirical evaluation of our
  technique shows an improved granularity of comparison.  It lowered the number
  of more precise invariants when comparing analysis techniques, thus, limiting
  the prevalent \emph{carry-over} effects. Moreover, it removed undecidable
  invariants and lowered the number of incomparable invariants when comparing
  two incomparable relational abstract domains.

\end{abstract}



\section{Introduction}\label{sec:intro}%

Various value-based static analysis techniques express computed program
invariants as a logical formula over program variables.  For example,
\acl{AI}~\cite{cousot-1977-abstr-inter} uses abstract domains such as
Zones~\cite{mine-2001-new-numer} and
Octagons~\cite{mine-2006-octag-abstr-domain} to describe an invariant as a set
of linear integer inequalities in a restricted format.  Other techniques such
as symbolic execution~\cite{king-1976-symbol-execut} and predicate analysis
combined with a symbolic component~\cite{sherman-2015-exploit-domain} do the
same, only using a general linear integer arithmetic format.  These invariants
are then used for program
verification~\cite{blanchet-2003-static-analyz,zhu-2018-a-data}, program
optimization~\cite{abate-2021-an-exten,katz-1978-progr-optim}, and for software
development tasks.

Static analysis developers rarely use a computed invariant by itself, but
rather compare them to determine effects of new algorithms or abstract domain
choices on the invariant precision.  For example, to evaluate tuning analyzer
parameters, static analysis researchers compare invariant values
\(\mathcal{I}\) and \(\til{\mathcal{I}}\) from the original and tuned analyzer runs,
respectively.  If an invariant becomes more precise, we conclude that the new
technique or a domain choice results in a more precise analysis.  For
relational domains one can use queries to an \acs{SMT} solver, such as
Z3~\cite{moura-2008-z3}, to determine which invariant is more precise by
checking their implication relations.

However, to objectively measure such effects in a computed invariant after
statement \(s\), \(\mathcal{I}_s\), we need to compare only the part of
\(\mathcal{I}_s\) affected by the transfer function of \(s\), \(\tau_s\).  This way, if
\(\til{\mathcal{I}}\) has already been more precise than \(\mathcal{I}\) before
\(s\) and \(\tau_s\) has not changed the relevant facts, then the comparison
should disregard the \emph{carry-over} precision improvement in
\(\til{\mathcal{I}}_s\).




The comparison of two relational invariants \(\mathcal{I}\) and
\(\til{\mathcal{I}}\) involves two steps: (1) identifying a changed component of each
invariant at a given statement and (2) performing minimal comparison between
the changed components of \(\mathcal{I}\) and \(\til{\mathcal{I}}\).  In our previous
work~\cite{ballou-2023-ident-minim} we addressed step (1) for the Zones domain
where using \acf{DFA} information, we developed efficient algorithms that find
a minimally changed set of inequalities in a Zone invariant.


In this work we target step (2), assuming that an abstract domain has some
means to perform step (1) using either elementary or sophisticated algorithms.
Thus, the contributions of this paper include: \textbf{(a)} development and
analysis of a minimal comparison algorithm for relational abstract domains and
\textbf{(b)} investigating its effect on comparisons between different widening
techniques for Zones domain as well as comparison between Zones and
incomparable Predicate domains with a relational component.

The rest of the paper is organized as follows.  In Section~\ref{sec:backgrd},
we provide the background, context, and motivation for our work.  In
Section~\ref{sec:approach}, we describe our fixed-point algorithm.  In
Section~\ref{sec:exper}, we explain our experimental setup and evaluation, and
in Section~\ref{sec:results}, we examine the results of our experiments.  We
connect this work with previous research in Section~\ref{sec:related}.
Finally, we conclude and discuss future work in Section~\ref{sec:concl}.



\section{Background and Motivation}\label{sec:backgrd}%

We refer to an invariant and the corresponding abstract domain as relational if
it is expressed as a conjunction of formulas over program variables,
\textit{e.g.,} a set of linear integer inequalities.  We first explain the
concept of the minimal change for an invariant and then explain challenges of
comparing two relational domains, and sketch how our proposed approach works.

\subsection{Minimal changes in relational abstract domains}

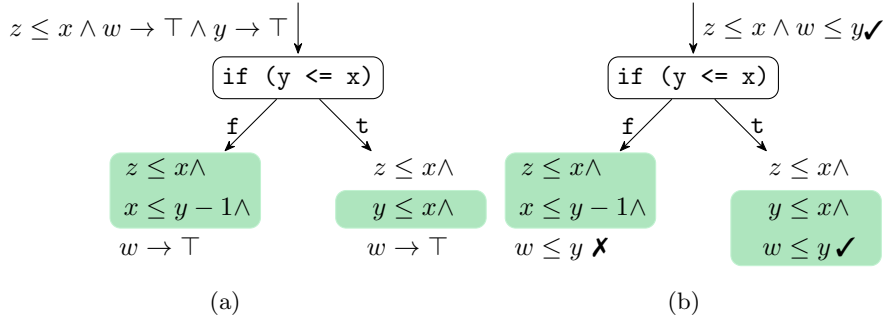
\begin{figure}[t]
  \centering
  \begin{subfigure}[b]{0.49\linewidth}
    \centering
    \begin{tikzpicture}[>={Stealth[round,sep]}]
      \draw [color=greenbg0, fill=greenbg1, rounded corners] (-2.5,-1.0) rectangle (-0.5,-2.0);
      \draw [color=greenbg0, fill=greenbg1, rounded corners] (0.5,-1.5) rectangle (2.5,-2.0);
      \coordinate (entry) at (0, 1);
      \coordinate (false) at (-1, -1);
      \coordinate (true) at (1, -1);
      \node[draw, rectangle, rounded corners] (branch) at (0, 0) {\texttt{if (y <= x)}};
      \node[text width=3cm] (f) at (-1.5, -1.5) {\begin{align*}
                                                   z &\le x \land \\[0pt]
                                                   x &\le y - 1 \land \\[0pt]
                                                   w &\to \top
                                                 \end{align*}};
      \node[text width=3cm] (t) at (1.5, -1.5) {\begin{align*}z &\le x \land \\
                                                 y &\le x \land \\
                                                 w &\to \top
                                               \end{align*}};

      \draw[->] (entry) to[left] node{\(z \le x \land w \to \top \land y \to \top\)} (branch);
      \draw[->] (branch) edge[left] node{\texttt{f}} (false);
      \draw[->] (branch) edge[right] node{\texttt{t}} (true);
    \end{tikzpicture}
    \caption{}%
    \label{fig:origSA}
  \end{subfigure}
  \begin{subfigure}[b]{0.49\linewidth}
    \centering
    \begin{tikzpicture}[>={Stealth[round,sep]}]
      \draw [color=greenbg0, fill=greenbg1, rounded corners] (-2.5,-1.0) rectangle (-0.5,-2.0);
      \draw [color=greenbg0, fill=greenbg1, rounded corners] (0.5,-1.5) rectangle (2.5,-2.5);
      \coordinate (entry) at (0, 1);
      \coordinate (false) at (-1, -1);
      \coordinate (true) at (1, -1);
      \node[draw, rectangle, rounded corners] (branch) at (0, 0) {\texttt{if (y <= x)}};
      \node[text width=3cm] (f) at (-1.5, -1.5) {\begin{align*}
                                                   z &\le x \land \\
                                                   x &\le y - 1 \land \\
                                                   w &\le y
                                                 \end{align*}};
      \node[text width=3cm] (t) at (1.5, -1.5) {\begin{align*}
                                                z &\le x \land \\
                                                y &\le x \land \\
                                                w &\le y
                                              \end{align*}};
      \draw[->] (entry) to[right] node (incoming) {\(z \le x \land w \le y\)} (branch);
      \draw[->] (branch) edge[left] node{\texttt{f}} (false);
      \draw[->] (branch) edge[right] node{\texttt{t}} (true);
      \node at (incoming.east) {\cmark};
      \node at ([xshift=0.75em,yshift=1em]f.south) {\xmark};
      \node at ([xshift=1.5em,yshift=1em]t.south) {\cmark};
    \end{tikzpicture}
    \caption{}%
    \label{fig:imprSA}
  \end{subfigure}
  \caption{Original Static Analysis (a) and Improved Static Analysis (b)}%
  \label{fig:example}
\end{figure}

Consider the relational invariants computed by a \acl{DFA} framework using the
Zones abstract domain as shown in Figure~\ref{fig:origSA}.  Let us assume the
analyzed code has four program variables: \(w\), \(x\), \(y\), and \(z\).
Here, the incoming flow to the conditional statement has the following
invariant: \(\mathcal{I}_{in} = z \le x \land w \to \top \land y \to \top\).  That is, variables
\(w\) and \(y\) are unbounded while \(x\) and \(z\) are bounded by a \(\le\)
relation.  The transfer function of the true branch adds \(y \le x\) inequality,
thus making \(y\) bounded.  This results in
\(\mathcal{I}_t = z \le x \land y \le x \land w \to \top\) invariant.  Similarly, the invariant
for the false branch becomes \(\mathcal{I}_f = z \le x \land x \le y - 1 \land w \to \top\).

Even though \(\mathcal{I}_f\) and \(\mathcal{I}_t\) are new invariants, they inherit two unchanged
inequalities \(z \le x\) and \(w \to \top\) from \(\mathcal{I}_{in}\).  This suggests that some
part of a previously computed invariants have not changed by the transfer
function of the conditional statement.  Thus, if for some application
\(\mathcal{I}_{in}\) is more precise because of \(z \le x\) and remains more precise in
\(\mathcal{I}_t\) because of the same inequality, such \emph{carry-over} precision
results should be disregarded.

Previous work determining minimal changes in a relational abstract domain
approach~\cite{ballou-2023-ident-minim} addresses this problem by identifying
the part the invariant affected by the statement's transfer function.  For
example, the minimal change algorithm for Zones~\cite{ballou-2023-ident-minim}
can compute the minimal sub-formula given the potentially changed variables
\(x\) and \(y\).  Specifically, the algorithm identifies only the \(y \le x\)
part of \(\mathcal{I}_t\) having changed from \(\mathcal{I}_{in}\).  Likewise for
\(\mathcal{I}_f\), the algorithm identifies two inequalities: \(z \le x\) and
\(x \le y - 1\) as the changed portion of the invariant.

The minimal change algorithm can be sophisticated and accurately compute the
changed part of the invariants, or can be over-approximating, and in the worst
case return the entire invariant.  In our previous work we developed an
efficient collection of such algorithms for the Zones abstract domain.  In this
work, we assume that a relational domain has an invariant change method
\(\Delta\) implemented, which takes as input an invariant and a set of updated
variables and returns a portion of \(\mathcal{I}\), \textit{e.g.,} in our example
\(\Delta(\mathcal{I}_t, \{x,y\}) = y \le x\).  The green shaded regions of an invariant in the
Figures~\ref{fig:origSA} and~\ref{fig:imprSA} indicates the changed part of the
state.

\subsection{Comparing relational domains}

Now consider invariants in Figure~\ref{fig:imprSA} computed for the same code
fragment, but using an improved algorithm.  This algorithm is able to compute
additional information for
\(\til{\mathcal{I}}_{in} = z \le x \land w \le y\), which is more precise than
\(\mathcal{I}_{in}\) since \(\til{\mathcal{I}}_{in}\) constrains the values of
\(w\) and \(y\).  The checkmark symbol, \cmark{}, by \(\til{\mathcal{I}}\) in
Figure~\ref{fig:imprSA} indicates an increased precision comparing to the
corresponding invariants \(\mathcal{I}\) in Figure~\ref{fig:origSA}.

When we compare using the entirety of the invariants instead of simply the
changed portion of the invariants for the false branch, the result would be
that \(\til{\mathcal{I}_f}\) is more precise than \(\mathcal{I}_t\).  Thus, simply applying
\(\Delta\) for both invariants can filter out erroneous, \emph{carry-over}
improvements, which we annotate with the \xmark{} symbol.

In the case of the true branch, the set of variables in their changed portion
of the invariant are the same.  However, this is not always the case, which we
can see on the false branches.  There,
\(\Delta(\mathcal{I}_f,\{x,y\}) = y \le x\), but
\(\Delta(\til{\mathcal{I}}_f,\{x,y\}) = y \le x \land w \le y\) has an extra variable
\(w\).  In order to make a sound comparison, we need to conjoin \(w \to \top\) with
the result of \(\Delta(\mathcal{I}_f,\{x,y\})\).  The challenge here is to identify the
smallest necessary additions to the changed portions of the invariants to
perform a sound comparison.

In the next section we present our proposed approach that addressees this
problem by developing a fixed-point algorithm that, in each iteration,
discovers a minimal set of inequalities (modulo \(\Delta\)) in one invariant that is
adequate for comparison with the changed part of the other invariant.



\section{Approach}\label{sec:approach}%

In this section, we explain the theoretical basis for our approach to minimally
compare relational invariants via logical entailment.  We start by defining the
problem, and then we present our algorithm that solves it.  At the end, we
perform the analysis of the proposed algorithm.



\subsection{Problem definition}

We define the problem in a context of a \acs{DFA} framework, where the
framework provides a set of updated variables \(dv\) that resulted in a new
invariant \(\mathcal{I}\).  An abstract domain for \(\mathcal{I}\) has a function
\(\Delta\) implemented, which returns a portion of \(\mathcal{I}\) that is changed by
\(dv\).  In the worst case \(\Delta(\mathcal{I}, dv) = \mathcal{I}\), \textit{i.e.,} the entire
invariant has been affected.  In the best case
\(\Delta(\mathcal{I}, dv) = \emptyset\), \textit{i.e.,} nothing has changed.  We also introduce a
function \(V\) that returns the set of variables used in \(\mathcal{I}\).  For example,
we use it to define the following property: \(V(\Delta(\mathcal{I}, dv)) \subseteq V(\mathcal{I})\).

Let \(\mathcal{I}_1\) and \(\mathcal{I}_2\) be two relational invariants, and let
\(dv_1\) and \(dv_2\) be their corresponding sets of updated variables.  Then the
problem of finding a minimal changed part of two invariants reduces to finding
a common minimal updated set of variables \(S\) such that

\begin{equation}\label{probDef}
S = V(\Delta(\mathcal{I}_1, S)) = V(\Delta(\mathcal{I}_2, S))
\end{equation}

A minimal solution for such recursive definitions is commonly obtained by a
fixed point iteration algorithm with initial values \(S_0\) set to the smallest
set, which in our case is \(S_0 = dv_1 \cup dv_2\).

\subsection{Finding a common changed variable set}

\begin{figure}[t]
  \centering
  \begin{algorithm}[H]
    \begin{algorithmic}[1]
      \Require{\(V(\mathcal{I}_1) = V(\mathcal{I}_2) \land V(\Delta(\mathcal{I}_1, dv_1)) \subseteq V(\mathcal{I}_1) \land V(\Delta(\mathcal{I}_2, dv_2)) \subseteq V(\mathcal{I}_2)\)}
      \Ensure{\(S_1 = S_2 \subseteq V(\mathcal{I}_1)\)}
      \Function{CommonVarSet}{$dv_1$, $dv_2$, $\mathcal{I}_1$, $\mathcal{I}_2$}
        \State{\(S_1 \gets \) \Call{V($\Delta$}{$\mathcal{I}_1$, $dv_1$})}\label{alg:union-initial-1}
        \State{\(S_2 \gets \) \Call{V($\Delta$}{$\mathcal{I}_2$, $dv_2$})}\label{alg:union-initial-2}
        \While{\(S_1 \neq S_2\)}\label{alg:union-loop}
          \If{\(S_1 \supset S_2\)}\label{alg:union-s1-sup-s2}
            \State{\(dv_2 \gets S_1 \setminus S_2\)}\label{alg:union-6}
            \State{\(S_2 \gets S_2 \cup \) \Call{$V(\Delta$}{$\mathcal{I}_2$, $dv_2$})}\label{alg:union-7}
          \ElsIf{\(S_2 \supset S_1\)}\label{alg:union-s2-sup-s1}
            \State{\(dv_1 \gets S_2 \setminus S_1\)}\label{alg:union-9}
            \State{\(S_1 \gets S_1 \cup \) \Call{V($\Delta$}{$\mathcal{I}_1$, $dv_1$})}\label{alg:union-10}
          \Else{}\label{alg:union-default}
            \State{\(dv_1 \gets S_2 \setminus S_1\)}\label{alg:union-12}
            \State{\(dv_2 \gets S_1 \setminus S_2\)}\label{alg:union-13}
            \State{\(S_1 \gets S_1 \cup \) \Call{V($\Delta$}{$\mathcal{I}_1$, $dv_1$})}\label{alg:union-14}
            \State{\(S_2 \gets S_2 \cup \) \Call{V($\Delta$}{$\mathcal{I}_2$, $dv_2$})}\label{alg:union-15}
          \EndIf{}
        \EndWhile{}
        \State{\Return \(S_1\)}\label{alg:union-return}
      \EndFunction{}
    \end{algorithmic}
    \caption{Common minimal changed variable set}%
    \label{alg:fixed-point-union}
  \end{algorithm}
  \vspace{-20pt}
\end{figure}

Algorithm~\ref{alg:fixed-point-union} shows the pseudocode of the optimized
fixed point computation algorithm to solve Equation~\ref{probDef}.  The
algorithm takes as arguments, the updated variables for each domain, \(dv_1\)
and \(dv_2\), two invariants to compare, \(\mathcal{I}_1\) and
\(\mathcal{I}_2\).  It requires basic conditions for its correctness: both invariants are
described over the same set of variables and \(\Delta\) does not introduce any new
variables.  The output is the solution for Equation~\ref{probDef}.


The algorithm first computes the initial changed variable sets, \(S_1\) and
\(S_2\) for each invariant, lines~\ref{alg:union-initial-1}
and~\ref{alg:union-initial-2}, affected by the updated variables \(dv_1\) and
\(dv_2\), respectively

At line~\ref{alg:union-loop}, the algorithm compares the two sets and if they
are not equal, \textit{i.e.,} the fixed point has not been reached, the
algorithm enters the main iteration loop.  Inside the body of the loop, the
algorithm first tests whether one set of variables is a proper superset of the
other, lines~\ref{alg:union-s1-sup-s2} and~\ref{alg:union-s2-sup-s1}.

If one of the sets is a proper superset, it only augments the smaller set as
done on lines~\ref{alg:union-6}--\ref{alg:union-7} and
lines~\ref{alg:union-9}--\ref{alg:union-10}.  For example, if
\(S_1 \supset S_2\), \(S_2\) is augmented by the variables which are not already in
\(S_2\).

Afterwards, a new updated variable set is computed from the set difference of
\(S_1\) and \(S_2\), line~\ref{alg:union-6}.  Then, the changed variable set is
computed as the union between the existing set \(S_2\) and the newly computed
minimum variables, line~\ref{alg:union-7}.  Similar computations are done for
the case when \(S_2 \supset S_1\), lines~\ref{alg:union-9}--\ref{alg:union-10}.

Finally, when the changed variable sets are incomparable---
line~\ref{alg:union-default}--- then both changed variable sets are recomputed in
a similar fashion as described in lines~\ref{alg:union-12}--\ref{alg:union-15}.
Upon the loop's termination, \textit{i.e.,} when \(S_1 = S_2\), the algorithm
returns one of the dependent sets, line~\ref{alg:union-return}.

To demonstrate how Algorithm~\ref{alg:fixed-point-union} compares two
invariants, consider the invariants on the true branch from our example in
Figure~\ref{fig:imprSA}.  There,
\(\mathcal{I}_1 = z \le x \land y \le x \land w \to \top\) and
\(\mathcal{I}_2 = z \le x \land y \le x \land w \le y\).  The updated variables are
\(dv_1 = \{x, y\}\) and \(dv_2 = \{x, y\}\).

The algorithm computes \(\{x, y\}\) for \(S_1\) and \(\{w, x, y\}\) for
\(S_2\).  Since \(S_2\) is a proper superset of \(S_1\), we recompute \(S_1\),
lines~\ref{alg:union-9} and~\ref{alg:union-10}.  Specifically, \(dv_1\) becomes
\(\{w\}\).  \(S_1\) is then recomputed:
\(S_1 = S_1 \cup V(\Delta(\mathcal{I}_1, dv_1))\), which results in
\(S_1 = \{x, y\} \cup \{w\} = \{w, x, y\}\).  At this point, \(S_1 = S_2\),
terminating the loop, and the algorithm returns the set \(S_1= \{w, x, y\}\).
Then, an \acs{SMT} solver can be used to compare logical relations of
\(\Delta(\mathcal{I}_1, S_1)\) and \(\Delta(\mathcal{I}_1, S_1)\), for example, using implication relations.
Or, in case of Zones, one can use its custom equivalence
engine~\cite{mine-2001-new-numer}.

As mentioned, under worst-case conditions,
Algorithm~\ref{alg:fixed-point-union} returns the entire set of variables.  In
other words, it devolves into a full invariant comparison.  This can happen if
the variables within the invariant are tightly coupled with all other
variables.  Another situation which can cause a worst-case comparison is when
an abstract domain has an ineffective \(\Delta\) function, which performs a basic
dependency analysis such as
slicing~\cite{ballou-2023-ident-minim,visser-2012-green}.

Below we present termination and complexity analysis for
Algorithm~\ref{alg:fixed-point-union}.  We start with a proof sketch of
termination.

\begin{proof}
  First, we begin with the following assumptions: the variable projections for
  both domains are equivalent, \textit{i.e.,}
  \(V(\mathcal{I}_1) = V(\mathcal{I}_2)\); and we assume the invariant minimization functions for
  each domain yield a subset of the variable projections, that is,
  \(\Delta(\mathcal{I}_1, dv_1) \subseteq V(\mathcal{I}_1)\), and similarly for \(\mathcal{I}_2\).

  At each iteration, the union of variables over the minimization function is
  always increasing by at least one variable in either \(S_1\) or \(S_2\).
  Therefore, within a finite number of iterations \(S_1\) and \(S_2\) reach
  fixed point, which is bounded by \(V(\mathcal{I}_1) = V(\mathcal{I}_2)\) condition.  Thus,
  Algorithm~\ref{alg:fixed-point-union} terminates.  \qed{}
\end{proof}

The complexity of Algorithm~\ref{alg:fixed-point-union} is
\(O(N) \cdot (C_{\Delta_1} + C_{\Delta_2})\), where \(N\) is the number of variables in the
program under analysis and $C_{\Delta_i}$ is the complexity of the invariant
minimization function for the corresponding domain.  In the worst-case, at each
iteration the sets \(S_1\) and \(S_2\) augmented by a single variable from
\(\Delta\) computations.  Overall, the time-complexity of
Algorithm~\ref{alg:fixed-point-union} depends on the number of variables and
the complexity of the \(\Delta\) functions of the abstract domains.



\section{Methodology}\label{sec:exper}%

To determine the effectiveness of the proposed algorithm, we use it to compare
invariants produced by different techniques and by different abstract domains
on the same program.  For each subject program, each analysis outputs
invariants after each statement.  Over the corpus of programs, we compute
\(6564\) total invariants.  We store the invariants as logical formulas in
\acs{SMT-LIB} format.  We run analyses on two relational domains, Zones and
Relational Predicates~\cite{sherman-2015-exploit-domain}, and compare the
results of a standard Zones analysis to advanced Zones analyses, and Zones
analysis to Relational Predicates analysis.

The goal of the empirical evaluation is to answer the following research
questions:

\begin{description}
\item[\textbf{RQ1}]{Does our technique affect the invariant comparison between
    different analysis techniques for the same abstract domain?}
\item[\textbf{RQ2}]{Does our technique affect the invariant comparison between
    two different relational domains?}
\item[\textbf{RQ3}]{How effective and efficient is
    Algorithm~\ref{alg:fixed-point-union} on real-world invariant comparisons?}
\end{description}

We consider different analysis techniques over the Zones domain to measure the
precision gained by various advanced techniques.  We consider the iteration
parameter before widening.  We also consider the widening method employed,
which ensures termination for Zones analysis.

We then compare the most precise Zones technique to Relational
Predicates~\cite{sherman-2015-exploit-domain}, two incomparable domains.  Our
previous work~\cite{ballou-2023-ident-minim} has shown the benefit of minimally
comparing incomparable domains to demonstrate realized precision.  However, in
this case, we extend the invariants of the Predicates domain with a symbolic
relational component.

For Relational Predicates, the minimization function is a selection based
solely on notions of variable reachability, \textit{e.g.,} variable dependence,
but it might not be minimal because of the generality of inequalities used in
the relational part.  We also computed minimization over Relational Predicates
using a purely connected component concept, similar to the technique by Visser
\textit{et~al.}~\cite{visser-2012-green}, however, reachable performed
marginally better.

We use the \acf{MN} minimization function from our previous
work~\cite{ballou-2023-ident-minim} for Zones which provides the smallest
invariant partition given a set of changed variables.  This minimization
algorithm considers the semantics of the formulas under the changed variables.
Using these semantics, it selects the minimal dependent substate from the
logical formula representing the invariant.

\subsubsection{Subject programs}

Our subject programs consist of \(192\) Java methods from previous research on
the Predicates domain~\cite{sherman-2015-exploit-domain}.  These methods were
extracted from a wide range of real-world, open-source projects and have a high
number of integer operations.  The subject programs range from \(1\) to
\(1993\) Jimple instructions, a three address intermediate representation.  The
average branch count for the methods is \(6\) (\(\sigma = 11\)), with one method
containing a maximal \(56\) branches.  A plurality of our subject methods,
\(81\), contain at least one loop, with one method containing \(12\) loops.

\subsubsection{Experimental platform}

We execute each of the analyses on a cluster of CentOS 7 GNU/Linux compute
nodes, running Linux version \texttt{3.10.0-1160.76.1}, each equipped with an
Intel Xeon Gold 6252 and \SI{192}{\giga\byte} of system memory.  We use an
existing \acs{DFA} static analysis
tool~\cite{ballou-2022-increm-trans,sherman-2015-exploit-domain} implemented in
the Java programming language.  The analysis framework uses
Soot~\cite{web-soot-oss,vallee-rai-1999-soot-java} version \texttt{4.2.1}.
Similarly, we use Z3~\cite{moura-2008-z3}, version \texttt{4.8.17} with Java
bindings to compare \acs{SMT} expressions for the abstract domain states.
Finally, we use Java version 11 to execute the analyses, providing the
following \acs{JVM} options: \texttt{-Xms4g}, \texttt{-XX:+UseG1GC},
\texttt{-XX:+UseStringDeduplication}, and \texttt{-XX:+UseNUMA}.

\subsubsection{Implementation}

We modified an existing \acs{DFA} framework such that the Zones analysis
outputs its entire invariant for each program point.  Each invariant is further
reduced using a redundant inequality reduction technique proposed by Larsen
\textit{et~al.}~\cite{larsen-1997-effic-verif}.  For all domains, unbounded
variables are set to top, \(\top\), and excluded from the output expression. This
further simplifies the formulas.  After entailment, we use Z3, using the
\acf{LIA} theory for Zones to Zones comparisons and the \acf{NIA} theory for
Zones to Relational Predicates comparisons, to decide model behavior of each
domain.

\subsubsection{Evaluations}

In total, we perform \emph{three} different invariant comparisons, summarized
in the following list:

\begin{description}
\item[\(Z \preceq Z_{k=5}\)|]{Zones using standard widening after two iterations and
    Zones widening after five iterations.}
\item[\(Z \preceq Z_{ths}\)|]{Zones with standard widening and Zones with threshold
    widening.}
\item[\(Z_{ths} \prec\succ P\)|]{Zones with threshold widening and Relational
    Predicates.}
\end{description}

In all instances of Zones sans \(Z_{k=5}\), widening happens after \emph{two}
iterations over widening nodes.  We use a generic set of thresholds for Zones
based on powers of 10: \(\{0, 1, 10, 100, 1000\}\).  Using a tuned set of
thresholds for each program would yield better results.

We use a generic disjoint domain for the basis of the Relational Predicates,
based on Collberg \textit{et~al.'s}~\cite{collberg-2007-empir-study} study of
numerical constants in Java Programs.  Specifically, the predicate domain used
in this study consists of the following set of disjoint elements:
\(\{(-\infty, -5]\), \((-5, -2]\), \(-1\), \(0\), \(1\), \([2, 5)\), \([5, +\infty)\}\).



\section{Evaluation Results and Discussions}\label{sec:results}%

In this section, we present the results of our experiments and discuss their
implications to the research questions posed in the previous section.

\subsection{Technique Comparisons}

\begin{table}[t]
  \centering
  \begin{minipage}{0.45\linewidth}\centering
    \begin{tabular}{|l|c|c|}
      \hline
      \bf{Comparison} & \(Z \equiv Z_{k=5}\) & \(Z \prec Z_{k=5}\) \\[0pt]
      \hline
      Full & 6555 & 9 \\[0pt]
      \hline
      Minimal & 6562 & 2 \\[0pt]
      \hline
    \end{tabular}
    \caption{Zones \(k=2\) widening compared to Zones \(k=5\) widening}%
    \label{tab:zones-zones-k}
  \end{minipage}
  \hfill
  \begin{minipage}{0.45\linewidth}\centering
    \begin{tabular}{|l|c|c|}
      \hline
      \bf{Comparison} & \(Z \equiv Z_{ths}\) & \(Z \prec Z_{ths}\)\\[0pt]
      \hline
      Full & 6519 & 45 \\[0pt]
      \hline
      Minimal & 6545 & 19 \\[0pt]
      \hline
    \end{tabular}
    \caption{Zones compared to Zones with Threshold Widening}%
    \label{tab:zones-zones-with-thresholds}
  \end{minipage}
\end{table}

To answer \textbf{RQ1}, we consider the comparisons of different techniques
using the Zones abstract domain.  Since different techniques using the same
domain create a partial ordering of their respective precision, we need only
consider equivalent and less precise outcomes.  To ensure correctness of our
implementation, we ensured that there were no other precision outcomes.

Table~\ref{tab:zones-zones-k} shows the breakdown of invariants computed by
standard widening after \emph{two} iterations and standard widening after
\emph{five} iterations.  Comparing invariants using the entire invariant,
deferred widening produces \emph{nine} more precise invariants.  However, when
using our minimized comparison technique, the slim advantage reduces to
\emph{two} invariants.

Table~\ref{tab:zones-zones-with-thresholds} shows the breakdown of invariants
between standard widening after two iterations and threshold widening after two
iterations.  Here, we see the largest gain in precision.  Using the entire
invariant to compare, threshold widening computes \(45\) more precise
invariants.  Again, however, the precision gain is cut by more than \(50\%\)
when using minimal comparisons.  The choice of thresholds could improve the
precision, but for best results, the set of thresholds needs to be tailored
specifically to each program.

As we can see between \(Z \preceq Z_{k=5}\) and \(Z \preceq Z_{ths}\), our comparison
technique lowers the number of more precise invariants, thus eliminating the
\emph{carry-over} precision instances.  That is, our technique lowers the
number of more precise invariants advanced techniques compute.  However, in
doing so, our technique presents a more granular image of the realized
precision gain advanced techniques offer.

\subsection{Zones versus Relational Predicates}

\begin{table}[t]
  \centering
  \begin{tabular}{|l|c|c|c|c|c|c|}
    \hline
    \bf{Comparison} & \(Z_{ths} \equiv P\) & \(Z_{ths} \prec P\) & \(Z_{ths} \succ P\) & \(Z_{ths} \prec\succ P\) & \(Z_{ths}\,?\,P\)\\[0pt]
    \hline
    Full & 1227 & 3173 & 196 & 1947 & 21 \\[0pt]
    \hline
    Minimal & 3675 & 2353 & 248 & 288 & 0\\[0pt]
    \hline
  \end{tabular}
  \caption{Zones with Threshold Widening compared to Relational Predicates}%
  \label{tab:zones-predicates}
\end{table}

Table~\ref{tab:zones-predicates} shows the precision breakdown of Zones with
threshold widening compared to Relational Predicates.  Given that Zones and
Predicates are inherently incomparable domains, we must consider all precision
comparison categories.  With the full invariant compairsions, Relational
Predicates are more precise than Zones in about \(50\%\) of the invariants.
The next largest category of invariants is incomparable, \(\prec\succ\), which accounts
for \(30\%\) of invariants.  Here, Zones and Predicates are complementary,
neither more nor less precise than the other.  Zones and Predicates are
equivalent in \(19\%\) of all invariants, and Zones are more precise in about
\(3\%\) of all invariants.  Finally, using the full invariant, \(21\) of the
program points between the relation between two invariants could not be
established by Z3 since it returned \texttt{UNKNOWN}.

Our technique eliminates the undecidable results.  Moreover, it dramatically
reduces the number of incomparable invariants-- only \(4\%\) of invariants
remain incomparable.  Similar to \emph{carry-over} precision, incomparable
invariants arise when one domain computes a more precise invariant for one
variable, and the other domain computes a more precise invariant for another,
unrelated variable at a later program point.  Considering the entire invariant
results in incomparable precision.  However, by comparing only the relevant,
changed variables, our technique largely disentangles the imprecision in the
comparison.

The equivalent invariant category is the next largest affected category, where
more than half, \(56\%\), of computed invariants between Zones and Relational
Predicates become equivalent.  Relational Predicates lose \(13\%\) of more
precise invariants, and Zones gains about \(1\%\) of invariants which it
computes more precisely than Relational Predicates.

By comparing only the necessary variables at each program point, our technique
allows general, relational abstract domains to be compared without undecidable
results.  The reduction in incomparable invariants between two otherwise
difficult to compare domains provides a clearer precision performance picture
between the two domains.

\subsection{Iterations and variable reductions}

\begin{figure}[t]
  \centering
  \input{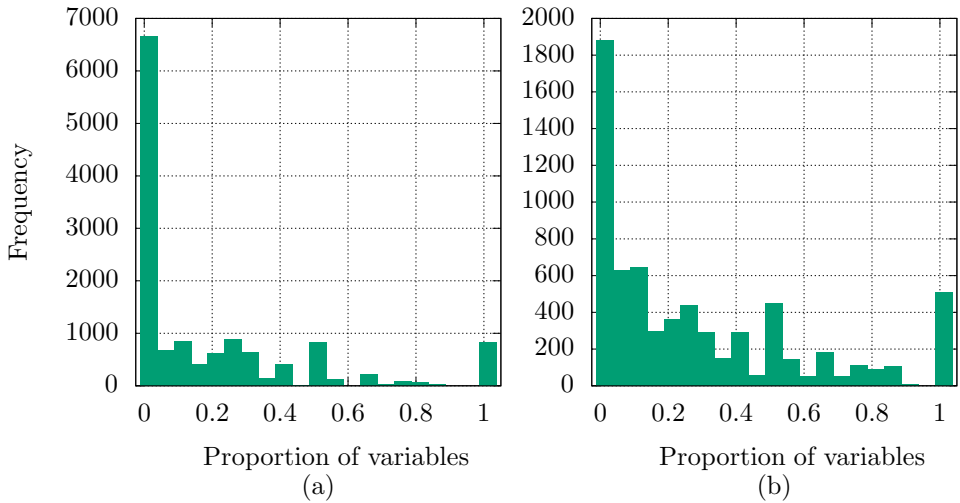}
  \caption{Frequency plot of proportion of variables selected by
    Algorithm~\ref{alg:fixed-point-union} which are necessary for comparing two
    invariants.  (a) represents the frequencies of proportions when comparing
    techniques using Zones.  (b) represents the frequencies of proportions when
    comparing Zones to Relational Predicates.}%
  \label{fig:variable-proportions}
\end{figure}

To determine if Algorithm~\ref{alg:fixed-point-union} is efficient,
\textbf{RQ3}, we use the iteration depth count to determine how many times the
algorithm iterates before it reaches a stable set of variables for comparison.
Over all instances of Zones comparisons, the iteration count was either
\emph{zero} or \emph{one}, with no outliers.  That is, either Zones computed
the same set of changed variables and the dependent set between two techniques
was immediately equivalent.  Or, the set of dependent variables is captured
with only a single extension, mostly to the Zones using standard widening,
\(Z\).

Comparing Zones to Relational Predicates, we see similar results.  The average
number of iterations is between \emph{zero} and \emph{one} iteration.  However,
we have several outliers at two iterations.  Instrumentation found \(12\)
instances of extreme outliers, \(11\) for \emph{three} iterations, and one
instance of \emph{four} iterations.  Furthermore, more variety exists in the
branches for Zones versus Relational Predicates.  Unlike comparing techniques
between Zones invariants, comparing Zones to a more general, relational formula
required more augmentation by each domain.

To evaluate effectiveness of Algorithm~\ref{alg:fixed-point-union},
\textbf{RQ3}, we consider the proportion of variables necessary for comparison.
We instrumented our algorithm to compute the proportion of variables it returns
after reaching a stable set, compared to the variable projection of the
incoming invariants.  We plot the frequency of proportions of variables
returned by Algorithm~\ref{alg:fixed-point-union} in
Figure~\ref{fig:variable-proportions}.  In
Figure~\ref{fig:variable-proportions}a, we plot variable reductions across all
comparisons of Zones: standard widening after two iterations versus standard
widening after five iterations and standard widening versus threshold widening.
Figure~\ref{fig:variable-proportions}b shows the variable reductions for Zones
with threshold widening versus Relational Predicates.  Considering a single bin
in Figure~\ref{fig:variable-proportions}, for example, \(0.1\), represents the
frequency where Algorithm~\ref{alg:fixed-point-union} needed only \(10\%\) of
the variables occurring in the original invariants to adequately compare the
two.

Shown in Figure~\ref{fig:variable-proportions}, the large frequencies in the
\(0\) bin shows our technique was able to remove all variables from the
invariants from comparison, eliminating the need to compare the two invariants.
Comparing advanced techniques utilizing Zones shows more than \(6500\)
instances, and about \(1850\) in Zones versus Relational Predicates.

Our technique reduces the number of variables necessary for comparison by
\(50\%\) or more in \(90\%\) of comparisons between techniques of Zones, and at
least by \(25\%\) in \(93\%\) of comparisons.  For Zones and Relational
Predicates, our technique reduces the necessary, relevant variables by \(50\%\)
or more in \(80\%\) of comparisons and by \(12\%\) in \(93\%\) of comparisons.
That is, in the majority of comparisons, our technique reduces the number of
variables necessary for comparing two relational domains or techniques.  The
quality of a domain's \(\Delta\) function affects the performance and effectiveness
of Algorithm~\ref{alg:fixed-point-union}.  We see only a few iterations in the
algorithm when comparing analysis techniques utilizing Zones since we used a
minimal \(\Delta\) function for Zones.  However, we see an increase in iterations
when comparing with a non-optimal \(\Delta\), as in Zones and Relational Predicates.
That is, the quality of \(\Delta\) can have an outsized impact on the practicality
of our technique.  However, given the preponderance of variable reductions and
low iteration counts over the corpus of methods and comparisons, we conclude
that the proposed algorithm is practical and effective.

\subsection{Discussion}

The evaluation results show our technique enables more precise comparison
between relational abstract domain invariants.  When comparing two techniques
using the same domain, our minimal comparison strategy precisely captures the
techniques' relative precision, disentangling accumulated, \emph{carry-over}
effects from realized precision gains.

While we do not have a proven state minimization function for Relational
Predicates, our technique still shows improvement when comparing incomparable
relational abstract domains.  Specifically, our comparison removes unknowns and
dramatically reduced incomparable invariants, which makes it easier to make
software engineering decisions.

The average iteration depth for Algorithm~\ref{alg:fixed-point-union} shows the
algorithm's efficiency and practicality.  Even when using an imprecise
minimization function for Relational Predicates, our technique only needed a
maximum of \emph{four} iterations to arrive at a stable set of common variables
for comparison.  Moreover, in the majority of comparisons,
Algorithm~\ref{alg:fixed-point-union} returned a significantly smaller
proportion of variables than the entirety of the variables in each invariant,
demonstrating the efficacy of the technique.



\section{Related Work}\label{sec:related}%

Our previous work~\cite{ballou-2023-ident-minim} found a set of algorithms for
efficiently computing \(\Delta\) for the Zones domain.  Using the algorithms, it
compared Zones to other non-relational domains, which in the context of
\acf{DFA} and this work, have trivial \(\Delta\) functions.  We extend the previous
work by considering comparisons between relational abstract domains,
abstracting the \(\Delta\) function for each domain.

Comparing the precision gain of new analysis techniques or comparing the
precision of newly proposed abstract domains is a common problem in the
literature.  Previous work in this area generally compare precision in one of
two ways.  One, the comparison is based on known \textit{a priori} program
properties on benchmark
programs~\cite{gange-2021-fresh-look,gurfinkel-2010-boxes,howe-2009-logah,laviron-2008-subpol,logozzo-2010-pentag}.
Two, the comparison is based on logical entailment of computed
invariants~\cite{howe-2009-logah,mine-2004-weakl-relat,sherman-2015-exploit-domain}.

To the best of our knowledge, this work represents one of the first studies
improving the granularity of precision characteristics for the latter
categorization of relational abstract precision comparisons.  We believe this
work would benefit existing work which compares relational abstract domains or
new analysis techniques using relational abstract domains.



\section{Conclusion and Future Work}\label{sec:concl}%

In this study, we defined the problem of minimally comparing relational
invariants, proposed an algorithm which solves the problem, and experimentally
evaluated whether the algorithm indeed solves the problem using real-world
programs.  Using our algorithm, we can remove the precision \emph{carry-over}
effects advanced analysis techniques introduce, providing clear precision
benefits for advanced techniques.  For example, the benefits of deferred
widening and threshold widening are smaller than anticipated.  Moreover, our
technique enables the comparison of relational abstract domains which are
otherwise difficult to compare directly.  Specifically, we see our technique
removed the \texttt{UNKNOWN} invariants and dramatically reduced the
incomparable invariants when comparing Zones to Relational Predicates.
Finally, Algorithm~\ref{alg:fixed-point-union}'s average iteration depth and
variable reduction demonstrate the algorithm's overall practicality and
usefulness when comparing analysis techniques and relational abstract domains.

\subsubsection{Future Work}

Developing a minimization function, \(\Delta\) for Relational Predicates would
enable a comprehensive, empirical study of the relative precision of
weakly-relational numerical abstract domains to Predicates.  Furthermore, we
believe the proposed technique of comparison can benefit adaptive analysis
techniques which selectively choose the appropriate abstract domain during
analysis.  Octagons~\cite{mine-2006-octag-abstr-domain} are not included in
this study because a minimization strategy for Octagons has not been developed.
However, this is an interesting avenue to pursue and we intend to use the
technique of this work to compare Zones to Octagons, which will empirically
quantify the precision gain of Octagons over Zones.



\nocite{tange_2022_6891516}
\bibliographystyle{splncs04}
\bibliography{bibfile}

\end{document}